\documentclass{article}

\usepackage{arxiv}

\usepackage[utf8]{inputenc} 
\usepackage[T1]{fontenc}    
\usepackage{booktabs}       
\usepackage{amsfonts}       
\usepackage{amsmath}
\usepackage{graphicx}
\usepackage{biblatex}
\usepackage[hidelinks]{hyperref} 

\title{Probabilistic and Alarm-Based Evaluation of a $b$-Value-Driven Deep Learning Earthquake Forecast}

\author{
 Jonas~Köhler\\
  Frankfurt Institute of Advanced Studies\\
  Frankfurt am Main\\
   \And
 Wei~Li \\
  Frankfurt Institute of Advanced Studies\\
  Frankfurt am Main\\
  \And
 Johannes~Faber \\
  Frankfurt Institute of Advanced Studies\\
  Frankfurt am Main\\
  and\\
  Institute of Theoretical Physics\\
  Goethe-University Frankfurt \\
  Frankfurt am Main \\
    \And
 Georg~Rümpker \\
  Frankfurt Institute of Advanced Studies\\
  Frankfurt am Main\\
  and\\
  Institute of Geosciences\\
  Goethe-University Frankfurt \\
  Frankfurt am Main \\
    \And
 Nishtha~Srivastava\\
  Frankfurt Institute of Advanced Studies\\
  Frankfurt am Main\\
  and\\
  Institute of Geosciences\\
  Goethe-University Frankfurt \\
  Frankfurt am Main\\
  \texttt{N.Srivastava@em.uni-frankfurt.de} \\
}

\newcommand{\Mw}{$\mathrm{M_{W}}$}
\newcommand{\MPerf}{\texttt{Model4.9}\ }

\newcommand{\bv}{$b$-value\ }
\newcommand{\bvs}{$b$-values\ }

\begin{document}
\maketitle
\begin{abstract}
We evaluate the forecasting performance of a deep learning model, originally introduced as a pattern-extraction framework, that operates on the spatiotemporal evolution of seismic $b$-values in a short-term forecasting context. Model output is rescaled to account for training on balanced datasets and evaluated relative to a spatial base-rate model using the Brier Skill Score (BSS). Absolute skill values are small, but mean BSS values are consistently positive, including at locations where \Mw $\geq 5$ earthquakes occurred during the test period, indicating information content beyond historical seismicity alone.
Alarm-based evaluation using Molchan diagrams shows elevated event capture rates at low alarm fractions ($5.88\%$ of events captured at $1\%$ area under alarm), indicating discrimination exceeding random and purely spatial reference models under constrained alarm conditions. Comparison with ETAS-derived triggered probabilities further reveals a weak positive correlation, suggesting partial sensitivity of the model output to seismic regimes characterized by enhanced clustering and recent activity, while remaining distinct from classical aftershock-based descriptions.
Together, these results indicate that spatiotemporal variations in $b$-values contain a persistent, though limited, signal relevant to probabilistic earthquake forecasting, yielding marginal but consistent improvements over baseline models across complementary evaluation frameworks.
\end{abstract}

\section{Introduction}
Earthquake forecasting remains a central challenge in seismology, not because seismicity lacks structure, but because that structure is subtle, scale-dependent, and embedded in highly imbalanced observational records. While long-term hazard assessments have achieved considerable maturity, short-term probabilistic forecasting-particularly at fine spatial and temporal resolution—continues to face fundamental limitations arising from data scarcity, non-stationarity, and the rarity of large events.

A wide range of empirical and statistical approaches has been developed to describe regularities in earthquake occurrence. One of the most enduring is the Gutenberg–Richter relation \cite{Gutenberg1944},
\begin{align}
    \log_{10} (N) = a-bM
\end{align}
which characterizes the relative abundance of small and large earthquakes through the \bv. Variations in this parameter have been linked to changes in stress conditions and faulting behavior, motivating its use in diverse contexts such as anomaly detection, rate modeling, and event classification \cite{Smith1981, Main1989, Smyth2011, Gulia2019}. In parallel, statistical point-process models, most prominently the Epidemic Type Aftershock Sequence (ETAS) framework \cite{Kagan1981, Kagan1987, Ogata1988, Ogata1998}, have provided a principled description of earthquake clustering driven by triggering and aftershock decay. While ETAS models are highly effective at capturing short-term rate increases following large events, they are not designed to forecast the occurrence of mainshocks themselves.

Recent advances in machine learning have opened new possibilities for extracting information from complex, high-dimensional geophysical data. In seismology, deep learning methods have proven successful in waveform-based tasks such as phase picking, event detection, and magnitude estimation, as well as in the analysis of geodetic observations \cite{LeCun2015, Zhu2018, Mousavi2020b, Chakraborty2022, Quinteros2023}. More recently, attention has shifted toward catalog-based forecasting approaches that aim to identify spatiotemporal patterns in seismicity itself. These include models based on binned seismic observables \cite{Fox2022}, auxiliary physical signals \cite{Saad2023}, neural point processes \cite{Stockmann2023}, and convolutional architectures operating on derived seismicity fields \cite{Koehler2023, Zhan2024, Zhan2025}.

Despite these developments, two issues remain central for practical forecasting applications. First, any learned signal must be evaluated against strong and well-understood baselines, such as spatial base-rate models or ETAS-derived expectations, to assess whether it captures information beyond historical seismicity and aftershock clustering. Second, forecast skill must be quantified using appropriate verification frameworks that account for the extreme class imbalance inherent in earthquake occurrence and allow meaningful comparison across models.

In this study, we investigate whether the information contained in the spatiotemporal evolution of \bvs can be exploited to produce probabilistic short-term forecasts that exceed simple baseline expectations. Building on a previously developed deep learning framework \cite{Koehler2026a}, we focus here not on methodological innovation, but on application-level evaluation. Using Japan as a test region, owing to its high seismicity rate and well-documented catalog completeness \cite{Wakita2013, Nanjo2010}, we assess daily forecasts of \Mw $\geq 5$ earthquakes on a $0.1^\circ \times 0.1^\circ$ grid.

Forecast performance is examined using complementary evaluation strategies, including proper scoring rules and alarm-based measures, and is interpreted in relation to ETAS-derived triggered probabilities. By situating the results within established seismological forecasting benchmarks, this work aims to clarify both the potential and the current limitations of deep learning approaches based on evolving seismicity indicators.

\section{Data and Base Model Summary}
\label{sec:data_base_model}

This section summarizes the data set and the base deep-learning model used throughout this study. The presentation is intentionally concise; methodological details, design choices, and validation of the base model are described in \cite{Koehler2026a}.

\subsection{Earthquake Catalog and Preprocessing}


We use the ISC earthquake catalog, restricted to the region $35^\circ - 46^\circ$\,N and $135^\circ - 146^\circ$\,E, covering the period from 1999-01-01 to 2023-12-31\footnote{Access date is 2025-12-17. At that time, the completeness dropped off in January 2024, hence our testing period.}. When moment magnitude \Mw  is not directly reported, the magnitudes are converted to \Mw using empirical relations from \cite{Scordilis2005}.

Data up to 2019-12-31 are used for model selection and training, while the subsequent four years (2020 to 2023) form the primary analysis period of this work. Importantly, the model is not static during this interval: training continues progressively (see \cite{Koehler2026a}), so that forecasts and scores always reflect the information available up to the respective evaluation time. During this time there are 125 events with \Mw $\geq 5$, which will be the targets of the forecasts here.

\subsection{Construction of $b$-Value Fields}

The input to the neural network consists of daily spatial fields of the Gutenberg-Richter \bv, computed on a $0.1^\circ \times 0.1^\circ$ grid. At each grid point, the \bv is estimated using the maximum-likelihood formulation \cite{Aki1965} and the $b$-positive method \cite{Elst2021}, based on earthquakes occurring within a spatial radius $r$ and a temporal lookback window $t$. 
The choice of $r = 0.6^\circ$ and $t = 365$ are taken from \cite{Koehler2026a} and conform to the \MPerf therein.

These daily \bv fields form a spatiotemporal data cube from which fixed-size blocks are extracted as model input.

\subsection{Base Model Architecture}

The base model follows a hybrid convolutional architecture combining spatial feature extraction with temporal sequence modeling. Spatial patterns in the \bv fields are captured by two-dimensional convolutional layers, while temporal evolution is modeled using one-dimensional temporal convolutional network (TCN) layers. The network is composed of repeated blocks of spatial convolution, temporal convolution, and normalization with nonlinear activation, followed by a final dense layer producing a single scalar output. The input to the model is a \bv block of size $512\times 32 \times 32$, corresponding to 512 days of \bv history and a $3.2^\circ\times 3.2^\circ$ spatial coverage. No architectural modifications are introduced in this paper; the exact layer configuration and design rationale are provided in \cite{Koehler2026a}.

\subsection{Progressive Training Scheme}

Training is performed using a progressive scheme in which the available data set grows monotonically over time. Starting on 2001-05-21 (512 days after 2000-01-01), the model is updated by sequentially adding non-overlapping 14-day periods to the training set. For each update, the newly added period contributes labeled samples, while earlier periods remain in the training pool. 

The training data are balanced at each stage: all samples associated with earthquakes of \Mw $ \geq 5$ are included, together with an equal number of samples drawn from periods without such events. Although short-term validation on subsequent unseen intervals is used during training, model selection based on this validation performance is extensively discussed in \cite{Koehler2026a} and is not central to the present study. In contrast to that work, the data here is from 1999 to 2023 (inclusive), so we have four years of testing data (2020 to 2023) for this work. Consequently, training also continues further than in the previous work.

\subsection{Base Model Output}

The base model produces a scalar output that can be interpreted as an anomaly score, reflecting whether a target region is associated with an elevated likelihood of an \Mw $\geq 5$ earthquake within the specified prediction window. Due to the enforced class balance during training, this output does not represent an absolute earthquake rate or a probability, and is therefore not directly suitable for forecasting applications. Converting this anomaly score into a calibrated rate-based forecast is the central focus of the remainder of this paper.

\section{Theoretical Background}
\label{sec:theory}

This section introduces the theoretical framework used to evaluate and interpret the model output after rescaling. The goal is to motivate the choice of probabilistic and alarm-based verification tools, rather than to provide an exhaustive review.

Probabilistic earthquake forecasts assign, for each space--time bin, a probability that at least one target event will occur. Evaluating such forecasts requires scoring rules that operate directly on probabilities. Proper scoring rules are constructed such that the expected score is optimized when the forecast probabilities coincide with the true probabilities of the underlying process, thereby discouraging hedging and systematic over- or underconfidence.

In the context of earthquake forecasting, verification is complicated by the rarity of target events and the strongly imbalanced nature of the binary outcome (event vs.\ no event). As discussed by \cite{Serafini2022}, some commonly used performance measures can behave counterintuitively in such low-probability environments, in particular by favoring forecasts that systematically overestimate or underestimate the true event probability. This issue is especially critical because the underlying probability $p^\ast$ is unknown in real forecasting experiments. Consequently, the choice of scoring rule must be made \emph{a priori}. Otherwise, improper scoring rules may on average assign a higher score to biased forecasts than to the underlying process itself. Proper scoring rules avoid this pathology by construction, as they guarantee that the highest expected score is achieved when forecast probabilities are closest to $p^\ast$. Following this framework, we adopt the Brier score as our primary measure for probabilistic forecast evaluation.

In addition to probabilistic scores, earthquake forecasts are often evaluated using alarm-based approaches that assess the trade-off between missed events and space-time volume under alarm. Molchan diagrams fall into this category and serve a different purpose: rather than evaluating probabilistic calibration, they characterize the performance of thresholded alarms and their associated error rates. In the following, we therefore use the Brier score to assess probabilistic forecast quality, and Molchan diagrams to provide a complementary, alarm-based perspective.

\paragraph{Brier Score}
\label{sec:brier}
One of the most widely used proper scoring rules for binary events is the Brier Score \cite{Brier1950}. For a set of $N$ forecasts, it is defined as
\begin{equation}
s_{\mathrm{BS}} = \frac{1}{N} \sum_{i=1}^{N} \left( p_i - r_i \right)^2 ,
\end{equation}
where $p_i$ denotes the forecast probability for the $i$-th space-time bin and $r_i \in \{0,1\}$ is the corresponding observed outcome. The Brier Score ranges from 0 (perfect agreement) to 1 (maximal disagreement), measuring the mean squared deviation between forecast probabilities and observations.

\paragraph{Brier Skill Score}
To assess forecast quality relative to a reference model, the Brier Skill Score (BSS) is commonly used:
\begin{equation}
s_{\mathrm{BSS}} = 1 - \frac{s_{\mathrm{BS}}^{\mathrm{model}}}{s_{\mathrm{BS}}^{\mathrm{baseline}}} .
\end{equation}
A BSS of zero indicates performance equal to the baseline, positive values indicate improvement, and negative values indicate degradation. The baseline can in principle be any other model, common uses would be a simple stationary model, or a direct competitor that the model is supposed to outperform.

The BSS is particularly suitable for rare-event forecasting, such as large earthquakes, because it evaluates the full probabilistic output rather than thresholded decisions. This avoids arbitrary alarm thresholds and allows improvements in both calibration and sharpness to be quantified, even when events occur infrequently.

\subsection{Molchan Diagrams}
\label{sec:molchan}

While probabilistic scores evaluate forecasts on an absolute scale, alarm-based methods focus on the trade-off between space-time coverage and event detection. These approaches are widely used in seismology and provide an alternative perspective on forecast performance.

Molchan diagrams \cite{Molchan1990, Zechar2008} evaluate forecasts by comparing the fraction of missed target events to the fraction of the space-time domain declared to be under alarm. For a given alarm threshold, regions exceeding that threshold are labeled as alarms, and the missed event rate is plotted against the corresponding alarm fraction. 

In contrast to receiver operating characteristic (ROC) curves, which display true positive and false positive rates, Molchan diagrams emphasize the cost of missed events relative to the imposed alarm volume. Common summary measures include the capture rate at fixed alarm fractions (typically 1\% or 5\%) and the area under the Molchan curve, where smaller values indicate better performance. Uncertainty in these measures can be estimated by bootstrapping over observed events.

Molchan diagrams are well suited to spatially and temporally extended forecasting problems, as they explicitly account for the fraction of the domain placed under alarm. However, they require binarization of the forecast through a threshold and therefore discard information contained in the full probability distribution. Their interpretation can also depend sensitively on how alarms are defined in space and time.

Because of these properties, Molchan diagrams complement Brier-type metrics. Probabilistic scores assess calibration and overall accuracy, while Molchan diagrams highlight the operational trade-off between alarm coverage and missed events. Using both perspectives allows a more complete assessment of forecast performance, particularly when transitioning from anomaly detection to rate-based earthquake forecasting.

\section{Forecast Construction via Rescaling}
The model in the current form as presented in \cite{Koehler2026a} outputs values that are not immediately useful for forecasting, as the imposed class balance during training strongly favors positive predictions. However, we want to provide a rescaling method to transform the original anomaly score into a usable probability or threshold output. To achieve this, we employ two different rescaling methods, one informed by experiment and one informed by theory.

All rescaling functions are calibrated exclusively on the training period and subsequently applied unchanged to the test period.
The results of these rescaling curves are shown in Figure~\ref{fig:BSS_scaling_heatmap}.

\subsection{Brier-score–based rate rescaling}
In the first approach, locations (in space and time) are grouped into bins according to the mean number of earthquakes $\overline{n_{\mathrm{eq}}}$ used for the local \bv estimation.\footnote{A single sample takes $512 \times 32 \times 32$ \bvs as the input. Each of these \bv calculations is based on a number of events, and this it the number we average to get $\overline{n_{\mathrm{eq}}}$.}
For each bin, the network outputs are rescaled by a multiplicative factor spanning $r \in [10^{-5}, 10^{-1}]$.
For each candidate scaling factor, we compute the Brier Skill Score (BSS) relative to the historical spatial bin-event rate.
This yields, for each $\overline{n_{\mathrm{eq}}}$ bin, a continuous range of scaling factors for which the BSS is positive or locally maximized.
We then fit a function of $\log \overline{n_{\mathrm{eq}}}$ to the midpoints of these ranges are regressed to obtain a smooth, monotonic rescaling function.
This empirical calibration aligns the average forecast rate with the observed event rate while preserving spatial discrimination.
The resulting scaling function is
\begin{align}
    s(n_{eq}) = 1.8 \times 10^{-5} \times e^{{n_{eq}}^{0.120}}
\end{align}

\subsection{Logit-based rescaling}
As an alternative, we apply rescaling in logit space, which directly accounts for the mismatch between the balanced training prior and the true, $\overline{n_{\mathrm{eq}}}$-dependent event rate.
Let
\begin{align}
\ell = \log\frac{p}{1-p}
\end{align}
denote the log-odds of the raw model output.
A change in the assumed class prior corresponds to an additive shift in log-odds,
\begin{align}
\ell' = \ell + \log\frac{\pi(\overline{n_{\mathrm{eq}}})}{1-\pi(\overline{n_{\mathrm{eq}}})},
\end{align}
where $\pi(\overline{n_{\mathrm{eq}}})$ denotes the expected event rate for a given $\overline{n_{\mathrm{eq}}}$ bin.

Assuming $\pi(\overline{n_{\mathrm{eq}}}) \propto \overline{n_{\mathrm{eq}}}^{-1}$ and $\pi(\max(\overline{n_{\mathrm{eq}}})) \ll 1$, this results in $1-\pi(\overline{n_{\mathrm{eq}}}) \approx 1$ and we obtain the rescaled probability:
\begin{align}
p' = \left(1 + e^{-\ell'}\right)^{-1}.
\end{align}

\begin{figure*}
    \includegraphics[width=\textwidth]{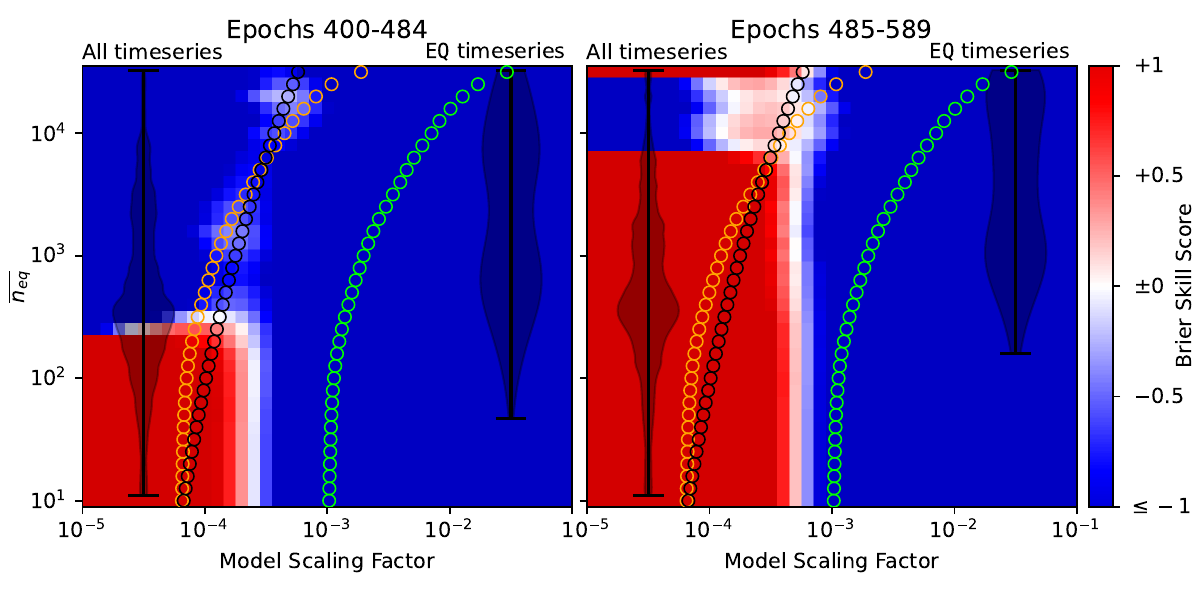}
    \caption[Model output rescaling]{
    Evaluation of model output rescaling using the Brier Skill Score (BSS) relative to the historical spatial bin-event rate.
    Each panel shows the BSS as a function of a multiplicative scaling factor and the mean number of earthquakes $\overline{n_{\mathrm{eq}}}$ used for local \bv\ estimation.
    The left panel corresponds to the training (calibration) epochs, and the right panel to the independent test epochs; BSS values are clipped at $-1$ for visualization.

Black circles indicate the empirical BSS-based rescaling obtained by fitting along the ridge of positive skill.
Green circles show the logit-based prior correction with the global event rate fixed from the training set.
Orange circles denote a logit-based rescaling with an additional offset calibrated to optimize BSS.
All rescaling functions are fitted on the training period and shown on the test period for diagnostic comparison only.

Violin plots illustrate the distribution of $\overline{n_{\mathrm{eq}}}$ for all evaluated space-time samples (left) and for samples culminating in an earthquake with \Mw $\geq 5$ (right) in each panel.

The red horizontal lines to the left are the result of less \Mw $\geq 5$ events happening in this bin in the test epochs.}
    \label{fig:BSS_scaling_heatmap}
\end{figure*}

We consider two variants of this logit-based rescaling.
In the first variant, the proportionality constant of $\pi(\overline{n_{\mathrm{eq}}})$ is fixed by matching the global event rate observed in the training set, yielding a prior-corrected but otherwise untuned transformation.
In the second variant, an additional constant offset in logit space is introduced and calibrated to maximize BSS on the training period.
This latter variant retains the functional dependence on $\overline{n_{\mathrm{eq}}}$ but allows for empirical adjustment of the absolute probability scale.
The logit-based rescaling function corresponding to a model output of $p=0.5$  are shown in Figure~\ref{fig:BSS_scaling_heatmap} for comparison with the empirical BSS-based calibration.

Logit-based rescaling preserves the rank ordering of model outputs while adjusting their absolute scale to reflect the rarity of target events, and avoids the loss of discriminative power associated with direct probability scaling in regions with large $\overline{n_{\mathrm{eq}}}$.

As expected, the empirical BSS-based rescaling aligns most closely with the regions of positive skill identified in the BSS diagnostics. For this reason, it is used for all subsequent analyses. The logit-based rescalings are shown for methodological comparison and to illustrate the effect of prior-based corrections.

\subsection{Molchan style Analysis}

We use this rescaled model output to create a Molchan diagram. As we need a discrete alarm-based prediction for the Molchan diagram, we define our threshold criterion as 
\begin{align}    
r_\mathrm{Model~Output} \times s(n_\mathrm{EQ}) \times k > p_\mathrm{Baseline~Model},
\end{align}
where $r_\mathrm{Model~Output}$ is the raw model output,  
$k$ is a single scalar threshold (scaled logarithmically from $10^{-2}$ to $10^{+10}$) to create the Molchan diagram and set the overall alarm rate, and 
$p_\mathrm{Baseline~Model}$ is the spatial baserate for \Mw $\geq$ 5 earthquake.
Confidence intervals are calculated using 10000 bootstrapping resamples.

\begin{figure}
    \includegraphics[width=0.5\linewidth]{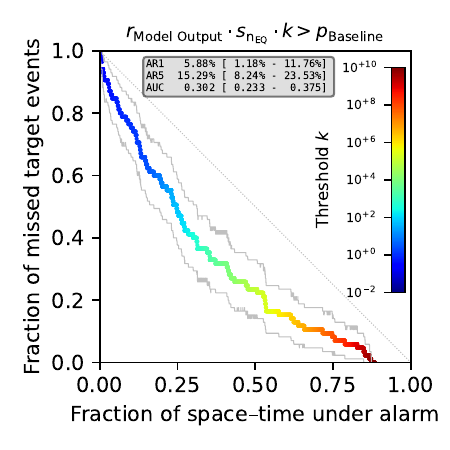}
    \caption{Molchan Diagram for \MPerf showing the alarm rates for $1\%$ and $5\%$ alarm rates and the AUC  with their respective $95\%$ confidence intervals.}
    \label{fig:molchan}
\end{figure}

\section{ETAS Comparison}
\label{sec:etas}
The Epidemic Type Aftershock Sequence (ETAS) is a model derived by \cite{Ogata1988, Ogata1998} based on earlier work from \cite{Kagan1981, Kagan1987} and is the most widely used earthquake (rate) forecasting model. The idea is that there is a group if independent background earthquakes happening at a constant rate. Each can trigger its own aftershocks, following the Gutenberg-Richter Law and the Omori-Utsu Law, and each aftershock can do the same, leading to an ``epidemic'' aftershock propagation.
Thus, at each point in time and space, there is an earthquake probability $\lambda$ based on all previous earthquakes and a background earthquake rate $\mu$. We assume a temporally and spatially uniform background rate \cite{Mizrahi2021}:
\begin{align*}
    \lambda (t, x, y) &= \mu + \sum\limits_{i:t_i < t} \nu_j(t-t_i, x-x_i, y-y_i, m_i)\\
    \nu(t, x, y, m) &= \frac{K_0}{\frac{(t + c)^{1+\omega}}{e^{-(t)/\tau}}} \cdot e^{\alpha(m - m_c)} \cdot \frac{1}{\left((x^2+y^2) + d \cdot e^{\gamma (m-m_c})\right)^{1+p}}
\end{align*}

To assess whether model behavior differs between background-like and cascade-dominated seismicity, we compare the model output to triggering probabilities derived from an independent ETAS declustering analysis. Rather than attempting a binary classification into ``mainshocks'' and ``aftershocks,'' we use the continuous ETAS triggering probability $P_{\text{triggered}}$ and background probability $P_{\text{background}} = 1-P_{\text{triggered}}$ assigned to each event, which quantifies the relative contribution of preceding seismicity to the conditional intensity at the event's occurrence time. In near-critical\footnote{Criticality here refers to the branching ratio, the number of aftershocks the average eathquake will have. The system becomes critical at a branching ratio of 1.} ETAS fits, such as those obtained here, $P_{\text{triggered}}$ is systematically high for most events, reflecting the dominance of triggering in the likelihood rather than a literal statement that most events are aftershocks. Importantly, the total expected number of triggered events remains finite, as given by the sum of triggering probabilities, even when individual probabilities are large.

\begin{figure}
    \includegraphics[width=\linewidth]{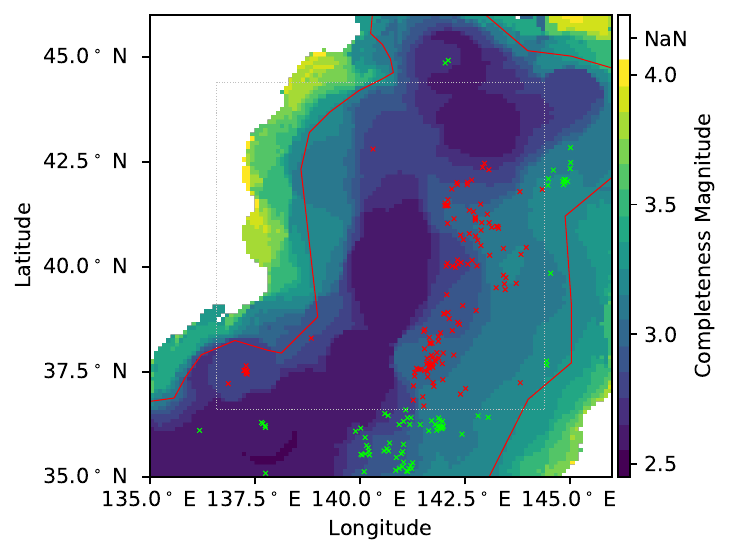}
    \caption{Completeness map including the ETAS region outline and the \Mw $\geq 5$ events. Completeness was determined using a $1^\circ$ radius around the center of each cell to collect the earthquakes, and then use the method in \cite{Godano2023} to determine $M_c$. As we do not visually inspect the curve for each cell, we determine $M_c$ as the smallest $m_{th}$ with $m_a(m_{th}) > \min(m_a) +0.3$. The red marks correspond to the events eligible to the model by being within $1.6^\circ$ degrees from the edge, a limit shown by the dotted gray line. The green marks correspond to \Mw $\geq 5$ events not included in the test data (no event from this region is the target in either training or testing).}
    \label{fig:etas_map}
\end{figure}

For this purpose, we use the ETAS implementation from \cite{Mizrahi2020, Mizrahi2021} for the inversion and determination of $P_{\text{background}}$ and $P_{\text{triggered}}$. The area as well as the accompanying magnitude completeness which was used to determine the area are shown in Figure~\ref{fig:etas_map}. 
As the magnitude completeness in the area is around 3.2 (determined using the method from \cite{Godano2023}) with a corresponding high number of events, we impose a cutoff magnitude of 3.5 for the ETAS inversion, leaving us with 51092 events in the area shown in Figure~\ref{fig:etas_map}. The inversion parameters are shown in Table~\ref{tab:etas_parameters}.

\begin{figure}
    \includegraphics[width=0.5\linewidth]{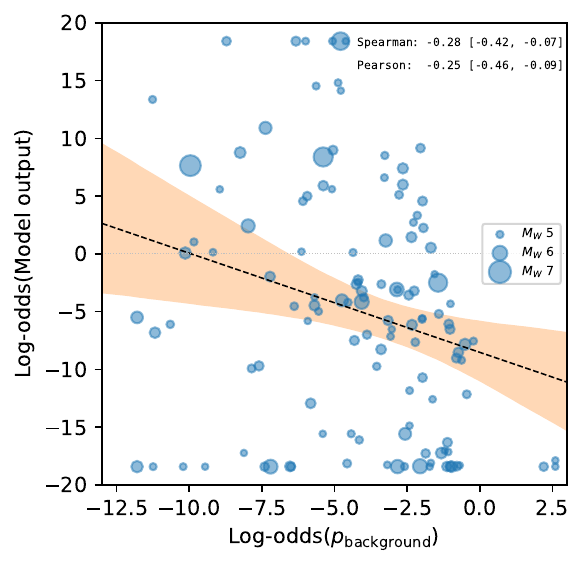}
    \caption{Scatter plot of model output against ETAS background probability. The scatter shows a small negative correlation between background probability and model output. The fit was created using a linear regression. Confidence intervals ($95\%$) on the noted correlations and the fit were calculated using 10000 bootstrapping resamples.}
    \label{fig:etas_scatter}
\end{figure}

Because the absolute values of $P_{\text{triggered}}$ depend on the fitted ETAS model (e.g., magnitude completeness and parameter trade-offs), they should not be interpreted as hard labels or compared across different declustering setups. Instead, their relative variation within a single fit provides a meaningful ordering of events according to their depth within the inferred triggering cascade. To visualize this relationship without imposing arbitrary thresholds or bins, we therefore examine the model output in logit transform $\ln \left( \frac{x}{1-x}\right)$, which expands the range of background-dominated events while retaining the continuous nature of the probabilities. 
This approach avoids common pitfalls in ETAS interpretation, such as equating high triggering probability with a discrete aftershock classification, and allows us to test whether the model output systematically varies with an event's relative position between background-like and cascade-dominated seismicity. The resulting Figure~\ref{fig:etas_scatter} shows a weak negative correlation between the logits of the model output and the logits of $P_\text{background}$, and a (therefore expected) positive correlation between the logarithm of the local productivity $\lambda_i$ and the logit of the model output.

\begin{figure}
    \includegraphics[width=\linewidth]{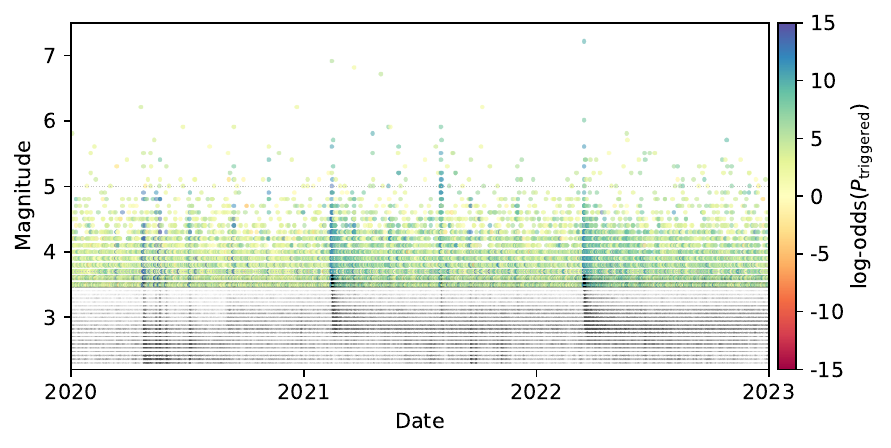}
    \caption{This figure shows the time-magnitude plot for the test set used in this work. The events are colored using the logits transformation of the ETAS derived probability that the earthquake is triggered. We show the earthquakes below the ETAS inversion threshold in 3.5 in black with low opacity.}
    \label{fig:etas_temporal}
\end{figure}

\begin{table}
    \centering
    \begin{tabular}{r|r}
    Parameter & Value \\
    \hline
    $\beta$ & $3.34$ \\
    $a$ & $2.32$ \\
    $\gamma$ & $2.63$\\
    $\log_{10}(c)$ & $-3.65$\\
    $\log_{10}(d)$ & $-0.66$\\
    $\log_{10}(k_0)$ & $-2.54$\\
    $\log_{10}(\mu)$ & $-5.57$\\
    $\log_{10}(\tau)$ & $2.88$\\
    $\omega$ & $-0.217$\\
    $\rho$ & $0.507$\\
    $b_r$ & $0.952$\\
    $\hat{n}$ & $2645$\\
    \end{tabular}
    \caption{ETAS parameters. The analysis takes $M_c = 3.5$. The inversion was performed using the code from \cite{Mizrahi2020, Mizrahi2021} and the parameter naming conventions are taken from there too.}
    \label{tab:etas_parameters}
\end{table}

\section{Discussion}
\subsection{Interpretation of Forecast Skill Relative to a Spatial Base Rate}
\begin{figure}
    \includegraphics[width=\linewidth]{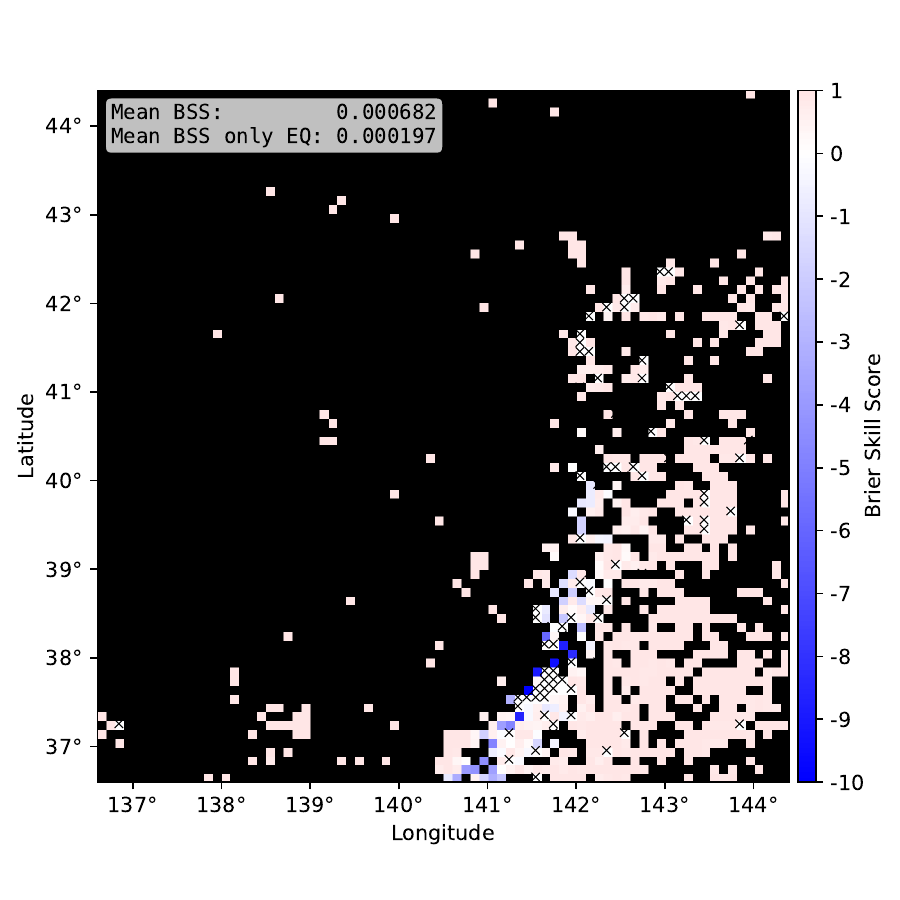}
    \caption{Brier Skill Score (BSS) map. BSS can only be computed for locations where the base rate score is non-zero; cells with no events during both training and testing periods are shown in black. Black crosses mark locations where earthquakes with  \Mw $\geq 5$ occurred in the test set. Mean BSS over all valid cells as well as only the cells which contain earthquakes during the test set are given in the top left.}
    \label{fig:BSS_map}
\end{figure}

The comparison against a spatial base rate model provides a stringent benchmark for assessing whether the model output contains information beyond historical seismicity alone. In this context, the observed Brier Skill Scores are small but consistently positive, both when averaged over all valid grid cells and when restricted to cells in which \Mw $\geq 5$ earthquakes occurred during the test period (see Figure~\ref{fig:BSS_map}). The mean BSS over the entire domain is $\overline{s}_{\mathrm{BSS}} = 0.000682$, while the mean over cells that hosted \Mw $\geq 5$ events during the test period remains slightly positive at $\overline{s}_{\mathrm{BSS}} = 0.000197$.

At the same time, an alarm-based evaluation using Molchan diagrams reveals a clearer separation from random performance. As shown in Figure~\ref{fig:molchan}, the model captures $5.88\%$ of \Mw $\geq 5$ earthquakes at a $1\%$ alarm fraction and $15.29\%$ at a $5\%$ alarm fraction.
 This indicates that the model is able to concentrate a nontrivial fraction of target events into a small fraction of space-time, demonstrating meaningful discriminatory power despite limited probabilistic calibration.

Such a divergence between Molchan-based performance and Brier Skill Scores is not unexpected for spatially resolved short-term earthquake forecasts of rare events. 
While the BSS penalizes probability miscalibration and base-rate mismatch, the Molchan diagram primarily reflects ranking and concentration ability. Taken together, the results indicate the presence of a weak but non-vanishing signal associated with the spatiotemporal evolution of \bvs while also highlighting substantial constraints on achieving strong probabilistic skill at these temporal and spatial scales.

\subsection{Effects of Training Data Imbalance and Major Sequences}
A major limitation affecting both training and evaluation is the dominance of the 2011 Tōhoku earthquake sequence in the available catalog. A large fraction of the positive training samples are associated with this single episode, leading to a strongly imbalanced representation of seismic regimes. While the implications of this imbalance for model training are discussed in detail in \cite{Koehler2026a}, it should be noted here that this concentration likely inflates early validation performance (in the years immediately after Tōhoku) and complicates the interpretation of aggregate skill measures later on. It should be noted, that we alreadey chose our benchmarking period (Epochs 400-484, 2016-09-17 to 2019-12-28) that short and that late to reduce the immediate effects of the Tōhoku earthquake sequence.

\subsection{Spatial Variability of Skill}
The spatial distribution of BSS values reveals marked regional variability. In particular, regions with comparatively low background seismicity tend to exhibit more consistently positive skill ($\bar{s_{BSS}} = 0.000685$), whereas highly active regions show mixed or near-neutral performance ($\bar{s_{BSS}} = 0.000197$). This pattern likely reflects a combination of effects, including differences in base rate, the influence of aftershock-dominated sequences, and the varying stability of \bv estimates as a function of event count ($n_{eq}$).

This spatial pattern is consistent with the Molchan analysis, which indicates that a disproportionate fraction of large events occurs within a small alarm fraction, even though the corresponding probabilities remain only weakly calibrated.

Importantly, positive skill is not confined to regions without large earthquakes; cells that hosted \Mw $\geq 5$ events during the test period still show a slightly positive mean BSS. This indicates that the model does not merely suppress probabilities in low-activity regions but retains limited discriminatory power even where large events occurred, a side effect that could be expected since the low-activity regions are more dominant and have a greater influence on the rescaling since they are more numerous (see Figure~\ref{fig:BSS_scaling_heatmap}). 

\subsection{Relation to ETAS-Based Seismicity Models}
In addition to the base rate comparison, the model output was examined in relation to ETAS-derived triggered probabilities. Although the two approaches are conceptually distinct, a weak but positive correlation is observed, with higher network output tending to coincide with events that ETAS classifies as having a higher triggered component.

This agreement should not be interpreted as equivalence between the two models or as evidence that the network reproduces ETAS-type triggering dynamics (the correlation is too weak to do that). Rather, it indicates that the spatiotemporal \bv patterns learned by the network are, to some extent, sensitive to seismic regimes characterized by elevated clustering and recent activity. At the same time, the substantial scatter in the relationship underscores that the model output cannot be reduced to an ETAS proxy and contains information that is at least partly distinct to a classical aftershock-based description.

\subsection{Relation to Previous $b$-Value Based Approaches}
Previous studies have reported systematic changes in \bvs preceding large earthquakes, using both descriptive and statistical approaches \cite{Smith1981, Main1989, Smyth2011, Gulia2019}. The present work differs in that it does not test a specific hypothesized pattern, but instead evaluates whether any information contained in the spatiotemporal evolution of \bvs can be exploited in an operational forecasting framework. The weak but positive skill observed here suggests that such information exists, while also indicating that using it in this current form is insufficient on its own to yield strong short-term probabilistic forecasts.

\section{Conclusion}
In this study, we evaluated the performance of the deep learning model based on the spatiotemporal evolution of \bvs originally presented in \cite{Koehler2026a} in an application-oriented forecasting context. The model output was rescaled to account for training on balanced datasets and assessed relative to a spatial base rate model using the Brier Skill Score. While the resulting BSS values are small in absolute terms, they are consistently positive on average, including at locations where \Mw $\geq 5$ earthquakes occurred during the test period, indicating the presence of a weak but non-vanishing signal beyond historical seismicity alone.

At the same time, an alarm-based evaluation using Molchan diagrams demonstrates a clearer separation from random performance. Under strict alarm constraints, the model captures a disproportionate fraction of target events, indicating meaningful spatial-temporal discrimination even when probabilistic calibration remains limited. Taken together, these results suggest that the model is more effective at concentrating earthquakes into a small fraction of space-time than at producing well-calibrated occurrence probabilities.

A complementary comparison with ETAS-derived background probabilities shows a weak negative correlation, suggesting that elevated model output partially coincides with seismic regimes characterized by increased clustering and recent activity. This consistency supports the physical plausibility of the learned patterns, while also highlighting that the model output cannot be reduced to classical aftershock-based descriptions.

At the same time, the results demonstrate clear limitations. Forecast skill over the short evaluation windows considered here remains modest, reflecting both the rarity of large earthquakes and the constraints imposed by the available catalog length. In particular, the dominance of the 2011 Tōhoku earthquake sequence likely introduces substantial imbalance into the training and evaluation data, complicating the interpretation of aggregate performance measures.

Overall, this work shows that information contained in the spatiotemporal evolution of \bvs can be exploited to produce forecasts that marginally outperform a spatial base rate in probabilistic terms, while exhibiting nontrivial discriminatory power under alarm-based evaluation. The results do not constitute a strong operational forecast, but rather provide an application-level validation that such signals persist under realistic evaluation conditions.

\section{Acknowledgements}
This research is supported by the “KI-Nachwuchswissenschaftlerinnen" -- grant SAI 01IS20059 by the Bundesministerium für Bildung und Forschung -- BMBF. The calculations were performed at the Frankfurt Institute for Advanced Studies’  GPU cluster, funded by BMBF for the project Seismologie und Artifizielle Intelligenz (SAI). We thank Megha Chakraborty, Darius Fenner, Dr.~Claudia Quinteros, Professor Geoffrey Fox, Dr.~Danijel Schorlemmer and Dr.~Kiran Thingbaijam for their helpful discussion. 

We acknowledge the help and advice from Prof.~Dr.~Horst Stoecker.
The research has made extensive use of PyTorch \cite{pyTorch2019}, numpy \cite{Numpy2020}, and matplotlib \cite{Matplotlib2007}.
Geographical maps were made with Natural Earth. Free vector and raster map data @ naturalearthdata.com.
The training was carried out on Nvidia A100 Tensor Core GPUs.

\printbibliography

\end{document}